\pdfoutput=1
\documentclass[aps,prl,twocolumn,preprintnumbers,superscriptaddress,10pt]{revtex4-2}

\usepackage[pdfborder={0 0 0},bookmarks=true]{hyperref}
\usepackage{graphicx,color,xcolor}
\definecolor{darkblue}{rgb}{0,0,.5}
\definecolor{darkgreen}{rgb}{0,0.5,.5}
\definecolor{darkyellow}{rgb}{0.5,0.5,0}
\definecolor{fhl}{rgb}{1,0,0}
\hypersetup{colorlinks=true, breaklinks=true, linkcolor=darkblue, menucolor=darkblue, urlcolor=darkblue, linktocpage=true}
\usepackage[english]{babel}
\usepackage{amssymb,amsfonts,amsmath}
\usepackage{bm}
\usepackage{bbm}
\usepackage{mathtools}
\usepackage{tikz}
\usetikzlibrary{arrows,shapes,backgrounds,calc,positioning,tikzmark,patterns,automata,decorations.markings}
\usepackage{verbatim}
\usepackage{picture}
\usepackage{cancel}
\usepackage{tabularx}
\usepackage{tensor}
\usepackage{fancyhdr}
\usepackage{sidecap}
\usepackage{scalerel}
\usepackage{relsize}
\usepackage[absolute,overlay]{textpos}
\usepackage[customcolors]{hf-tikz}
\usepackage{cals}
\usepackage{enumitem}
\setlist{nolistsep}
\usepackage{pifont}
\usepackage[export]{adjustbox}
\usepackage[most]{tcolorbox}
\usepackage{varwidth}
\tcbuselibrary{skins}

\newcommand{\be}{\begin{equation}}
	\newcommand{\ee}{\end{equation}}
\newcommand{\ba}{\begin{eqnarray}}
	\newcommand{\ea}{\end{eqnarray}}
\newcommand{\baa}{\begin{array}}
	\newcommand{\eaa}{\end{array}}

\newcommand{\tr}{{\text{tr}}}
\newcommand{\SEE}{S_{\text{EE}}}
\newcommand{\dSEE}{\partial_\ell\SEE}

\newcommand{\ie}{{\emph{i.e.}}}
\newcommand{\eg}{{\emph{e.g.}}}

\let\originalleft\left
\let\originalright\right
\renewcommand{\left}{\mathopen{}\mathclose\bgroup\originalleft}
\renewcommand{\right}{\aftergroup\egroup\originalright}

\newcommand{\e}{\operatorname{e}}

\newcommand{\SU}[1]{\operatorname{SU}(#1)}

\newcommand{\On}[1]{\operatorname{O}\left(#1\right)}
\newcommand{\Un}[1]{\operatorname{U}\left(#1\right)}

\newcommand{\of}[1]{\left(#1\right)}

\newcommand{\bof}[1]{\biggl(\bigg.#1\bigg.\biggr)}

\newcommand{\sof}[1]{\bigl(\big.#1\big.\bigr)}
\newcommand{\ssof}[1]{(#1)}
\newcommand{\fof}[1]{\left[#1\right]}

\newcommand{\ssfof}[1]{[#1]}
\newcommand{\cof}[1]{\left\{#1\right\}}

\newcommand{\bcof}[1]{\biggl\{\bigg.#1\bigg.\biggr\}}

\newcommand{\sscof}[1]{\{#1\}}

\newcommand{\ii}{\mathrm{i}}

\newcommand{\dd}{\mathrm{d}}

\newcommand{\partd}[2]{\frac{\partial #1}{\partial #2}}
\newcommand{\partdf}[3]{\left.\frac{\partial #1}{\partial #2}\right\vert_{#3}}

\newcommand{\ssorder}[1]{\mathcal{O}(#1)}

\newcommand{\ssabs}[1]{| #1|}

\renewcommand*\[{\begin{equation}}
\renewcommand*\]{\end{equation}}
\renewcommand*\hat[1]{\widehat{#1}}
\let\oldstackrel\stackrel
\renewcommand*\stackrel[2]{{\scriptstyle\oldstackrel{#1}{#2}}}
\definecolor{emphcol}{RGB}{0,0,0}
\let\oldemph\emph
\renewcommand*\emph[1]{\oldemph{\textcolor{emphcol}{#1}}}
\let\oldstackrel\stackrel
\renewcommand*\stackrel[2]{{\scriptstyle\oldstackrel{#1}{#2}}}

\newcommand{\ucases}[1]{\begin{cases}#1\end{cases}}

\newcommand*\getscale[1]{%
  \begingroup
    \pgfgettransformentries{\scaleA}{\scaleB}{\scaleC}{\scaleD}{\whatevs}{\whatevs}%
    \pgfmathsetmacro{#1}{sqrt(abs(\scaleA*\scaleD-\scaleB*\scaleC))}%
    \expandafter
  \endgroup
  \expandafter\edef\expandafter#1\expandafter{#1}%
}

\newcommand{\nrep}{\tilde{n}}

\begin{document}

\title{Thermal and chemical response from entanglement entropy}

\author{Niko Jokela}
\email{niko.jokela@helsinki.fi}
\affiliation{Department of Physics and Helsinki Institute of Physics,\\
P.O.~Box 64, FI-00014 University of Helsinki, Finland}

\author{Aatu Rajala}
\email{aatu.rajala@helsinki.fi}
\affiliation{Department of Physics and Helsinki Institute of Physics,\\
P.O.~Box 64, FI-00014 University of Helsinki, Finland}

\author{Tobias Rindlisbacher}
\email{tobias.rindlisbacher@helsinki.fi}
\affiliation{Department of Physics and Helsinki Institute of Physics,\\
P.O.~Box 64, FI-00014 University of Helsinki, Finland}

\begin{abstract}
We study entanglement entropy (EE) in interacting quantum field theories (QFTs) at finite density. We argue that, in the limit of large subregions, the derivative of EE with respect to the size of the entangling region approaches the thermal entropy density, independently of microscopic details. We make this relation explicit using slab-shaped subregions, where the limiting behavior can be directly identified. At finite chemical potential, we show that EE satisfies thermodynamic response relations, including a generalized Maxwell relation linking chemical potential and charge density. We provide strong nonperturbative evidence for these statements in the three-dimensional $\On{4}$ model, and conjecture that they are generic features of QFTs, establishing a two-way link between entanglement and thermodynamics that opens a route toward extracting the equation-of-state information from entanglement data.
\end{abstract}

\preprint{HIP-2026-4/TH}


\maketitle

Understanding the structure of interacting quantum field theories (QFTs) at finite density remains a central challenge in theoretical physics. In this regime, the presence of a chemical potential reshapes the many body state and its correlations, yet first principles access to thermodynamic and response properties is often limited.

Entanglement entropy (EE) provides a natural way to characterize quantum correlations in a basis-independent manner. In QFT, EE is also a subtle observable: it is ultraviolet (UV) divergent and depends on the short distance regulator, typically exhibiting an area law divergence controlled by correlations across the entangling surface~\cite{Srednicki:1993im,Eisert:2008ur,Casini:2009sr,Nishioka:2018khk}. For this reason, EE itself is not expected to be universal. The situation changes when one focuses on how EE varies as the entangling region is enlarged. In thermal states, and more generally when the linear size of the region is large compared to the correlation length, EE contains an extensive contribution proportional to the thermal entropy density $s$, while the UV sensitive area term is independent of the region size at leading order. This structure is explicit in soluble QFT settings and scaling arguments~\cite{Calabrese:2004eu,Swingle:2011np}. It motivates the expectation that suitable size derivatives of EE isolate bulk thermodynamic physics, even though EE itself remains UV sensitive.

In this Letter, we demonstrate these ideas explicitly in the three-dimensional interacting $\On{4}$ model, where R\'enyi entropies can be computed nonperturbatively even at finite density $n$ and used as estimators of EE. We argue on general grounds that in the limit of large subregions, the derivative of EE with respect to the size of the entangling region approaches the thermal entropy density. We exemplify the relation for slab-shaped subregions, where the limiting behavior can be identified explicitly. At finite chemical potential $\mu$, we further show that variations of EE satisfy thermodynamic response relations, including a generalized Maxwell relation linking chemical potential and charge density. Unlike the first law of entanglement~\cite{Bhattacharya:2012mi,Blanco:2013joa,Belin:2013uta}, which relates variations of EE to expectation values of the modular Hamiltonian for fixed regions, the relations studied here arise from varying the size of the entangling region itself and directly probe bulk thermodynamic structure.

Technical details and extended analyses are presented in a companion paper~\cite{Jokela:2026tkv}.

To establish the thermodynamic content of EE, we first consider the standard Landau free energy density
\begin{equation}
\omega_{L}\of{T,\mu}=\epsilon-T\,s-\mu\,n=-p\ ,\label{eq:landaufe}
 \ \dd\omega_{L}=-s\,\dd T - n\,\dd\mu\ ,
\end{equation}
where $\epsilon$ is the internal energy density. We define a dimensionless free energy density, $\omega\of{\beta,\mu}$, in terms of a grand canonical partition function $Z\of{\beta,V,\mu}$, 
\begin{equation}
\omega\of{\beta,\mu}=-\lim_{V\to\infty}\frac{1}{V}\log\of{Z\of{\beta,V,\mu}}\ ,\label{eq:latfe}
\end{equation}
where $\beta=1/T$ is the inverse temperature (Euclidean time extent) and $V=L^{d-1}$ is the $(d-1)$-dimensional, $L$-periodic spatial volume.
The dimensionless quantity $\omega\of{\beta,\mu}$ from \eqref{eq:latfe} relates to \eqref{eq:landaufe} via 
\begin{equation}
\omega\of{\beta,\mu}=\beta\,\omega_{L}\of{1/\beta, \mu}\ ,  \ \dd\omega=\frac{\omega+s}{\beta}\,\dd \beta - \beta\,n\,\dd\mu\ .\label{eq:latfediff}
\end{equation}
From \eqref{eq:latfediff}, the thermal entropy density is extracted as
\begin{equation}
s=\beta\,\partdf{\omega}{\beta}{\mu}-\omega  \ .\label{eq:thermsdensfromlatfe}
\end{equation}

Next, we define the EE associated with a spatial region $A$ in a volume $V$ as
\be
 \SEE(A) = -\tr\left(\rho_A\log \rho_A\right) \ ,
\ee 
where $\rho_A=\tr_{B}\of{\rho}$ is the reduced density matrix of the degrees of freedom supported in $A=V\setminus B$ and the density matrix $\rho$ of the full system corresponds to a pure state, \ie, $\rho^2=\rho$. While conceptually straightforward, the logarithm renders $\SEE$ difficult to compute in interacting QFTs, necessitating, \eg, replica techniques~~\cite{Calabrese:2004eu,Calabrese:2009qy}.

The replica method relates $\tr(\rho_A^r)$ to a ratio of Euclidean path integrals,
\begin{equation} \label{replica}
    \text{tr}(\rho^r_A)=\frac{\tilde Z(A,\beta,V,\mu,r)}{Z(\beta,V,\mu)^r}\ ,
\end{equation}
where $\tilde Z(A,\beta,V,\mu,r)$ is the path integral for the field theory in a geometry where the Euclidean time is $r\beta$-periodic over region $A$ and splits into $r$ copies of periodicity $\beta$ over region $B$. Quantities in the replicated geometry are denoted by a tilde. Parameter dependencies (\eg, $\beta,V,\mu$) are suppressed whenever unambiguous. The EE then follows as
\begin{eqnarray}
       \SEE(A) & = & -\lim_{r\rightarrow 1}\frac{\partial\log\text{tr}(\rho^r_A)}{\partial r}  \label{eq:EE} \\
       & = & 
         -\of{\lim_{r \rightarrow 1}\frac{\partial\log \tilde Z(A,r)}{\partial r}-\log Z}  \ , \nonumber  
\end{eqnarray}
which, in order for $\rho_A=\tr_{B}\of{\rho}$ to be the reduced density matrix of a pure state $\rho$, should be considered in the limit $\of{\beta\to\infty}$. However, it seems customary to refer to \eqref{eq:EE} as \emph{entanglement entropy} also at finite temperature despite thermal entropy contributions~\cite{Calabrese:2004eu,Calabrese:2009qy} and we will do so as well.

For simplicity, let us now assume that the entangling region $A$ is a slab of width $\ell$. The $\ell$-derivative of the EE reads
\begin{equation}
\partdf{\SEE\of{\ell}}{\ell}{\beta,V,\mu}
=-\lim_{r\to 1}\partd{}{r} \bof{\partd{\log \tilde Z(\ell,r)}{\ell}}\ ,\label{eq:seelsderiv}
\end{equation}
where we assumed that taking the limit $\of{r\to 1}$ commutes with taking the $\ell$-derivative.
We now use the argument presented in~\cite{Jokela:2023rba}, that for $\xi\ll\ell\ll L$, \ie, if the linear sizes of the entangling region $A$ and its complement $B$ are both much larger than the longest correlation length $\xi$ of the theory, then one has
\begin{multline}
-\lim_{\underset{\ell\ll L}{\ell,L\to\infty}}\frac{1}{V_{\perp}}\partdf{\log \tilde Z(\ell,r)}{\ell}{\beta,V,\mu,r}\\
=\omega\of{r\,\beta,\mu} - r\,\omega\of{\beta,\mu}\ ,\label{eq:felatlargelapprox}
\end{multline}
where $V_{\perp}=\partial_{\ell}\mathrm{Vol}\of{A}=L^{\of{d-2}}$.
By combining \eqref{eq:seelsderiv} with \eqref{eq:felatlargelapprox} and using \eqref{eq:thermsdensfromlatfe} we obtain
\begin{equation}
\lim_{\underset{\ell\ll L}{\ell,L\to\infty}}\frac{1}{V_{\perp}}\partdf{\SEE\of{\ell}}{\ell}{\beta,V,\mu} = s\of{T\of{\beta},\mu} \ , \label{eq:eederiveqtherms}
\end{equation}
and further using (\ref{eq:landaufe}) we recall
\begin{equation}
 s\of{T,\mu}
 =-\partdf{\omega_L}{T}{\mu} \ .\label{eq:entropydensfromlandaufe}
\end{equation}

Thus, the $\ell$-derivative of EE  equals (minus) the $T$-derivative of the Landau free energy. Note that in \eqref{eq:eederiveqtherms} the derivative can equivalently be understood as a derivative with respect to spatial size of $A$. More precisely, it corresponds to varying the extent of region $A$ while keeping the total system volume $V$ fixed, together with the shape and orientation of the entangling surface $\partial A$. In this sense, the derivative probes how the EE changes under an infinitesimal rigid displacement of $\partial A$, rather than under a deformation of its geometry.

Interestingly, this behavior (\ref{eq:eederiveqtherms}) appears to also persist in the deconfining phase~\cite{Nakagawa:2010kjk,Jokela:2023rba}, where previous studies in Yang--Mills theories demonstrated saturation at large $\ell$ and the expected temperature scaling, though a precise quantitative comparison to the thermal entropy is required. This regime was originally motivated by holography, where the relation follows by construction from the Ryu--Takayanagi prescription~\cite{Ryu:2006bv}.
More broadly, holography has suggested that $\dSEE$ acts as a probe of the finite correlation length~\cite{Jokela:2020wgs} at large-$N$, particularly in confining phases~\cite{Nishioka:2006gr,Klebanov:2007ws} where the number of degrees of freedom scales as $\ssorder{N^0}$ and $\dSEE$ is expected to drop rapidly at large $\ell$. 

The relation (\ref{eq:eederiveqtherms}) derived here is precise and nonperturbative, and is in fact most sharply realized in phases with finite $\xi$. The replica construction serves only as an intermediate computational device; the resulting relation between $\dSEE$ and thermodynamic quantities is independent of the replica formalism.

From the Maxwell relation $\of{\partial_{\mu}s}\vert_{T}=\of{\partial_{T}n}\vert_{\mu}$, which is a consequence of $\dd\dd\omega_L=0$, we obtain, after replacing $s$ by $V_{\perp}^{-1}\dSEE$ according to \eqref{eq:eederiveqtherms}, the relation
\begin{equation}
\frac{1}{V_\perp}\frac{\partial^2\SEE}{\partial\mu\,\partial\ell}\Bigg|_{\beta,V,\mu}=-\beta^2\,\partdf{n}{\beta}{\mu}\ \ , \ \  \xi\ll\ell\ll L \ .\label{eq:thermsdensmaxrel}
\end{equation}

Relations analogous to \eqref{eq:eederiveqtherms} and \eqref{eq:thermsdensmaxrel} can also be derived for the R\'enyi entropies of integer order $r\geq 2$,
\begin{align} \label{eq:renyi}
    H_r\of{\ell}&\equiv \frac{1}{1-r}\log\tr\of{\rho^r_A} \nonumber \\
    &=\frac{1}{1-r}\of{\log \tilde Z\of{\ell,r}-r\log Z}\ ,
\end{align}
for which one has:
\begin{equation}\label{eq:HrlogZrelation}
\frac{\partial H_r\of{\ell}}{\partial \ell} = \frac{1}{1-r}\frac{\partial\log \tilde{Z}\of{\ell,r}}{\partial \ell}\ .
\end{equation}
It then follows from \eqref{eq:felatlargelapprox} that
\begin{equation}
\lim_{\underset{\ell\ll L}{\ell,L\to\infty}}\frac{1}{V_{\perp}}\frac{\partial H_r\of{\ell}}{\partial \ell}=s_{r}\of{T\of{\beta},\mu}\ ,\label{eq:hromegaexpr}
\end{equation}
with
\begin{multline}
s_r\of{T,\mu}=-\frac{\omega_L\of{T,\mu}-\omega_L\of{T/r,\mu}}{T-T/r}\\
\equiv -\Delta_{T}^{r}\,\omega_L\of{T,\mu}\ ,\label{eq:stepscaleentropydens}
\end{multline}
where we used $\omega\of{\beta,\mu}=\beta\,\omega_L\of{1/\beta,\mu}$ from \eqref{eq:latfediff} and $T=1/\beta$, and introduced the \emph{step scaling derivative} with respect to $T$ for a scaling factor $r$. The latter yields a discrete approximation of the partial derivative with respect to $T$ in terms of a \emph{step scaling function} with scaling factor $r$. When extending $r$ to $\mathbb{R}_{\geq 1}$ one has 
\begin{equation}
\lim_{r\to 1} s_r\of{T,\mu}=s\of{T,\mu}\ ,\label{eq:stepscaleentropydenslimit}
\end{equation}
which is consistent with the fact that $\SEE\of{\ell}=\lim_{r\to 1} H_r\of{\ell}$, so that \eqref{eq:hromegaexpr} reduces to \eqref{eq:eederiveqtherms} in that limit.

By taking a derivative with respect to $\mu$ on both sides of \eqref{eq:hromegaexpr}, one obtains
\begin{equation}
\frac{1}{V_{\perp}}\frac{\partial^2 H_r}{\partial\mu\,\partial\ell}=\Delta_{T}^{r}\,n\of{T,\mu}|_{T=T\of{\beta}}\ ,\ \ \xi\ll\ell\ll L\ ,\label{eq:maxwell_rel_hr}
\end{equation}
where the right-hand side is again a discrete step scaling approximation to the right-hand side of \eqref{eq:thermsdensmaxrel} with scaling factor $r$, and one has
\begin{equation}\label{eq:nstepscalederiv}
\lim_{r\to 1}\Delta_{T}^{r}\,n\of{T,\mu} = \partdf{n}{T}{\mu}=-\beta^2\,\partdf{n}{\beta}{\mu}\ ,
\end{equation}
so that in the limit $r\to 1$ equation \eqref{eq:maxwell_rel_hr} reduces to \eqref{eq:thermsdensmaxrel}.

We now specialize from general considerations of entanglement and R\'enyi entropies in QFTs to the $\On{N}$ models used in our simulations. Despite their simplicity, $\On{N}$ models exhibit rich physics and provide a well controlled testing ground for new ideas. Crucially, they can be simulated directly at finite density on the lattice using worm algorithms~\cite{Prokofev:2001ddj}, unlike, \eg, quantum chromodynamics where one must rely on extrapolation techniques~\cite{Philipsen:2007rj,deForcrand:2009zkb}. Together with their modest computational cost, this makes $\On{N}$ models ideal for testing the relation \eqref{eq:thermsdensmaxrel} and its step scaling approximation \eqref{eq:maxwell_rel_hr}.

From now on, all quantities are expressed in lattice units, with the lattice spacing $a$ kept implicit.

The lattice action of a general $\On{N}$ model with sources $j$ and a chemical potential $\mu$ that couples to a conserved $\Un{1}$ charge, is
\begin{align}    
S\fof{\phi}=&\,\sum\limits_{x}\Bigl\{-\frac{\kappa}{2}\,\sum\limits_{\nu=1}^{d}(\phi_{x}\,e^{\mu\,\tau_{12}\,\delta_{\nu,d}}\,\phi_{x+\hat{\nu}} \nonumber \\
&+\phi_{x}\,e^{-\mu\,\tau_{12}\,\delta_{\nu,d}}\,\phi_{x-\hat{\nu}}\,)  \\
&+\,(\phi_x \cdot \phi_x)+\lambda\,((\phi_x \cdot \phi_x)-1)^{2}-j\cdot \phi_x\Bigr\}\ ,  \nonumber
\end{align}
where $\kappa$ is a hopping parameter, $\lambda$ a coupling, and $\tau_{12}$ is an $N \times N$ matrix with components ${\tau_{12}}\indices{^a_b}=\ii \of{\delta_{1,a}\delta_{2,b}-\delta_{1,b}\delta_{2,a}}$, $\forall\,a,b\in\cof{1,\ldots,N}$. 

At nonzero $\mu$, the action becomes complex, preventing importance sampling in terms of the original fields $\phi$. This sign problem can be avoided by reformulating the theory in terms of integer-valued dual \emph{flux variables}, which can be efficiently sampled using a worm algorithm \cite{Prokofev:2001ddj}. Several dual formulations and corresponding update schemes exist. We use the approach developed in \cite{Rindlisbacher:2015xku,Rindlisbacher:2016zht,Rindlisbacher:2017ysn}. For alternative dual representations and applications beyond the $\On{N}$ model, see \cite{Gattringer:2012df,Gattringer:2012ap,Endres:2006xu,Bruckmann:2015sua,Katz:2016azl} and \cite{Schmidt:2012uy,DelgadoMercado:2012tte,Bruckmann:2015sua,Rindlisbacher:2016cpj}, respectively.

In the chosen formalism, the partition function $Z=\int\mathcal{D}\phi\, \exp\ssof{-S\ssfof{\phi}}$, written in terms of dual variables, is~\cite{Jokela:2026tkv,Rindlisbacher:2017ysn,Rajala:2026mve}
\begin{multline} \label{eq:partition function}
Z=\sum\limits_{\cof{k,l,\chi,p,q,n}}\prod\limits_{x}\bcof{\delta\sof{p_x+\sum_{\nu}\of{k_{x,\nu}-k_{x-\hat{\nu},\nu}}}\\
\times\bof{\prod\limits_{i=3}^{N}\delta_{2}\of{L_{x}^{i}+M_{x}^{i}}}\,\e^{\mu\,k_{x,d}}\\
\times\bof{\prod\limits_{\nu=1}^{d} w_{\mathrm{l}}\of{L_{x,\nu}; \kappa}}\,w_{\mathrm{s}}\of{L_{x},M_{x}; \lambda, j}}\ ,
\end{multline}
where $k_{x,\nu}\in\mathbb{Z}$ counts the charged net flux from site $x$ to site $x+\hat{\nu}$ on the link $\of{x,\nu}$, $l_{x,\nu}\in\mathbb{N}_0$ the number of neutral pairs of charged particles, and $\chi_{x,\nu}^{\of{i}}\in\mathbb{N}_0$ the number of neutral particles of type $i$ moving along that link. The \emph{monomer numbers}, $p_x\in\mathbb{Z}$, $q_x\in\mathbb{N}_0$, and $n^{\of{i}}_x\in\mathbb{N}_0$ are the total charge, the number of neutral pairs of charge particles, and neutral particles of type $i$ at site $x$, respectively. Furthermore, \eqref{eq:partition function} makes use of the following abbreviations: $L_{x,\nu}=\sof{k_{x,\nu},l_{x,\nu},\chi_{x,\nu}^{\of{3}},\ldots,\chi_{x,\nu}^{\of{N}}}$ is a $N$-tuple describing the flux variable configuration on the link $\of{x,\nu}$, $L_x$ a $N$-tuple with components $L^{i}_x=\sum_{\nu=1}^{d}\ssof{\ssabs{L^{i}_{x,\nu}}+\ssabs{L^{i}_{x-\hat{\nu},\nu}}}$, counting the flux of type $i$ on links attached to site $x$, and $M_x=\sof{p_x,q_x,n_x^{\of{3}},\ldots,n_x^{\of{N}}}$ the $N£$-tuple describing the monomer configuration on site $x$. The functions $w_l\of{\cdot;\kappa}$ and $w_s\of{\cdot,\cdot;\lambda,j}$ are link and site weights, respectively. The remaining factors in \eqref{eq:partition function} are the on-site constraints, imposed by the discrete delta function $\delta\of{\cdot}$, 
enforcing local charge conservation, and the evenness constraint $\delta_2\of{\cdot}$, requiring the sum of the flux variables of a given type, adjacent to a site, and the corresponding monomer number on that site to sum up to an even number. These on-site constraints prohibit importance sampling by updating individual $k$, $\chi^{\of{i}}$, $p$, or $n^{\of{i}}$ variables, but make the system ideally suited for being updated with a worm algorithm. Details on the used algorithm can be found in~\cite{Rindlisbacher:2015xku,Rindlisbacher:2016zht,Rindlisbacher:2017ysn,Jokela:2026tkv}. The implementation supports both the linear and nonlinear versions of the $\On{N}$ model.

We now outline how $\dSEE$ can be measured nonperturbatively in lattice $\On{N}$ models with the replica method. Since we cannot take the limit $r\to 1$ in numerical computations, we approximate $\partial_r$ in \eqref{eq:EE} by a discrete forward derivative~\cite{Buividovich:2008kq,Rabenstein:2018bri}. This corresponds to estimating $\SEE$ with the 2nd R\'enyi entropy $H_2\of{\ell}$ (cf.~\eqref{eq:renyi}. We then need to approximate also the $\ell$-derivative of $H_2\of{\ell}$ with a finite difference, leading to
\begin{equation} \label{eq:dSEE_latt}
       \partdf{\SEE\of{\ell'}}{\ell'}{\mathrlap{\ell'=\ell+1/2}}\ \ \ \approx\ \partdf{H_2\of{\ell'}}{\ell'}{\mathrlap{\ell'=\ell+1/2}}\ \  = -\log\bof{\frac{\tilde{Z}\of{\ell+1,2}}{\tilde{Z}\of{\ell,2}}}\ . 
\end{equation}

In more than one spatial dimension, the distribution of configurations contributing to $\tilde Z(\ell+1,2)$ is likely to have very little overlap with the distribution of configurations contributing to $\tilde Z(\ell,2)$, since the actions of the two systems then differ for $\ssorder{V_{\perp}}\gg 1$ sites. This \emph{overlap problem} makes a direct evaluation of \eqref{eq:dSEE_latt} impractical. 

To overcome this overlap problem, we use the boundary-deformation method introduced in \cite{Rindlisbacher:2022bhe,Jokela:2023rba} for $\SU{N}$ gauge theories. The method allows the update algorithm to move back and forth along a sequence of updates that, piece by piece, deform the boundary $\partial A$ of the entangling region $A$ in an ordered manner to interpolate between the states of $\partial A$ in $\tilde Z(\ell,2)$ and $\tilde Z(\ell+1,2)$. By recording histograms representing the relative frequencies with which the algorithm visits the different states in this sequence of boundary deformed states, $\dSEE$ is obtained as the logarithm of the ratio between the histograms for the first and last states. The histograms are stored periodically, so that statistical uncertainties can be estimated using jackknife resampling.

Several alternative strategies have been proposed to address the poor overlap between ensembles at different $\ell$~\cite{Buividovich:2008kq,Nakagawa:2009jk,Nakagawa:2010kjk,Itou:2015cyu,Rabenstein:2018bri,Alba:2016bcp,Bulgarelli:2023ofi,Bulgarelli:2024onj,Bulgarelli:2024yrz,Bulgarelli:2025ewp}. 
 
For certain EE observables, the problem can be avoided altogether, \eg, when studying entanglement in the presence of static quark-antiquark pairs~\cite{Amorosso:2024leg,Amorosso:2024glf,Amorosso:2026mdo,Amorosso:2026zkj}. Tensor-network methods provide another independent route to entanglement measures~\cite{Coser:2013qda,Yang:2015rra,Bazavov:2017hzi,Cataldi:2023xki,Hayazaki:2025srr}.

It is worth mentioning that the on-site constraints in \eqref{eq:partition function} cause in principle even with the boundary deformation method a severe overlap problem: a local deformation of the entangling region boundary implies that the temporal boundary conditions over a spatial site change, causing $r=2$ temporal links to exchange their endpoints. The latter alters the incoming flux on these endpoints and likely causes defects, \ie, violations of the on-site constraints on these sites, resulting in a weight zero configuration. This overlap problem can be resolved by using worm updates to either remove defects caused by the change of temporal boundary conditions, or to manipulate the flux variable configurations before and after the change of temporal boundary conditions so that the formation of defects is prevented. Both strategies can be turned into efficient boundary update algorithms that respect detailed balance, as described in detail in the companion paper~\cite{Jokela:2026tkv}.
 
\begin{figure}[!htb]
    \centering
    \includegraphics[width=0.8\linewidth]{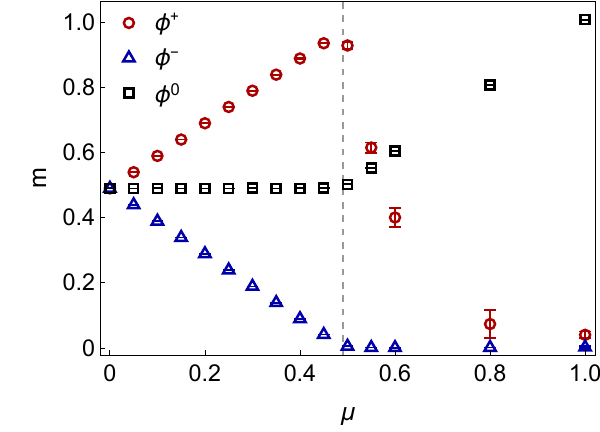}
    \caption{Mass spectrum 
    at $\kappa=1.2$, $j_3=0.2$ as a function of $\mu$. Note that the $\phi^+$ mass is only accurately determined up to the critical $\mu\approx 0.5$, since at finite density, $\phi^+$ has overlap with the vacuum and is no longer a well-defined particle state.}
    \label{fig:massspectrum}
\end{figure}

In our simulations we choose the hopping parameter $\kappa=1.2$, placing the system well inside the phase where the global $\On{4}$ symmetry is spontaneously broken to $\On{3}$. In this regime, the radial mode is heavy and effectively decouples, leaving three Goldstone modes, $\sscof{\phi^+, \phi^-, \phi^0}$, as the relevant low-energy degrees of freedom. Since relations \eqref{eq:felatlargelapprox} and \eqref{eq:thermsdensmaxrel} require a finite correlation length, we introduce a source $j_3=0.2$, which gives the Goldstone modes a mass $m_0\approx 0.5$ at $\mu=0$. 
For $\mu>0$ the masses of $\sscof{\phi^+, \phi^-, \phi^0}$ split, as shown in figure~\ref{fig:massspectrum} and the mass of the lightest mode, $\phi^{-}$, behaves as
\begin{equation}
    m^{-}\of{\mu}=\ucases{m_0-\mu\quad\text{if}\quad\mu<m_0\\0\quad\text{otherwise}}\ .
\end{equation}
The longest correlation length in the system is, therefore, given by $\xi_{\text{max}}\of{\mu}=1/m^{-}\of{\mu}$ and diverges as $\mu$ approaches the critical value $\mu_c=m_0\approx 0.5$.

We simulate systems with $r=2$ replicas on lattices of size $r\,N_t\,V$ with spatial lattice volume $V=N_x\,V_{\perp}$ and $V_{\perp}=N_s^{d-2}$ with $d=3$ and $N_s=12$, and furthermore $N_x=36$ and $N_t=5,\ldots,10$. The width $\ell$ of the entangling region is measured along the $x$-direction and is interpolated between $\ell=17$ and $18$, so that the derivative with respect to $\ell$, evaluated as a finite difference, is computed for $\ell=17.5$. Simulations of the unreplicated system are carried out on lattices of size $N_t\,N_s^{d-1}$, with $N_t=5,\ldots,20$.

To test the relation \eqref{eq:thermsdensmaxrel}, the charge density has to be defined on the lattice in terms of the dual variables. We will explicitly distinguish between the charge density of the unreplicated system,
\begin{equation} \label{eq:lattice_n}
    n=\frac{1}{N_t\,V}\frac{\partial \log Z}{\partial \mu}=\frac{\langle \sum_xk_{x,d}\rangle_{Z}}{N_t\,V} \ 
\end{equation}
and the one of the $r$-replica system,
\begin{equation} \label{eq:lattice_n_2}
    \nrep=\frac{1}{r\,N_t\,V}\frac{\partial \log \tilde Z}{\partial \mu}=\frac{\langle \sum_xk_{x,d}\rangle_{\tilde{Z}}}{r\,N_t\,V} \ .
\end{equation}

Before addressing relation \eqref{eq:thermsdensmaxrel}, we present a consistency test for our simulation algorithm by determining $\partial_{\ell}\partial_{\mu} \log \tilde Z$ for arbitrary values of $\mu$ and $\ell$ in two different ways, namely a) by using measurements of $\tilde{n}$ at given $\mu$ and two subsequent values of $\ell$ to compute a discrete $\ell$ derivative approximation for $\partial_{\ell}\tilde{n}$, and b) by computing $\partial_{\ell}H_2$ at given $\ell$ for chemical potential values $\mu\pm\Delta\mu$ to compute a discrete $\mu$-derivative approximation for $\partial_{\mu}\of{\partial_{\ell}H_2}$. Since $\partial_{\ell}H_2=-\partial_{\ell}\log \tilde Z$ (cf.~\eqref{eq:HrlogZrelation}), it follows that
\begin{equation} \label{eq:d2F}
    \frac{\partial^2 H_2}{\partial\mu\,\partial\ell} = -2\,N_t\,V \frac{\partial \nrep}{\partial\ell}\ .
\end{equation}

\begin{figure}[!htb]
    \centering
    \includegraphics[width=0.8\linewidth]{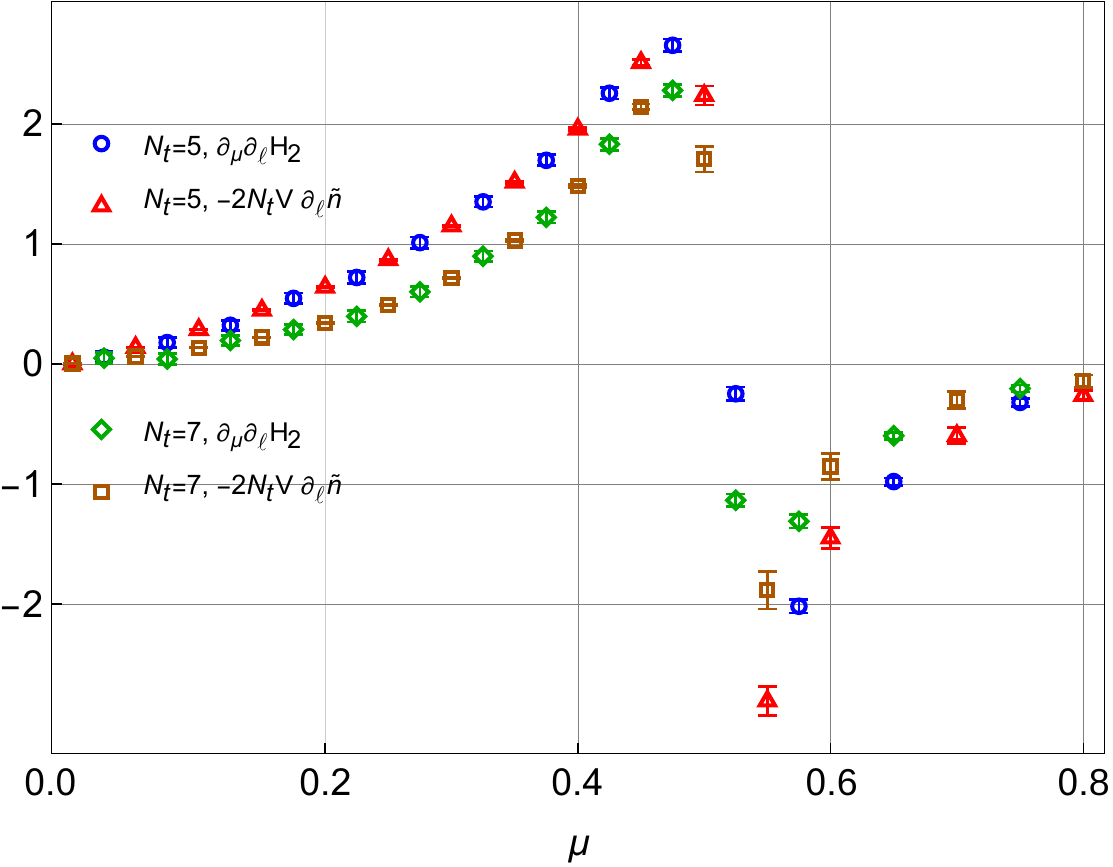}
    \caption{Comparison of results for $-\partial_{\mu}\partial_{\ell} \log \tilde Z$ as computed from $\partial_\mu\partial_{\ell} H_2$ and $-2\,N_t\,V\,\partial_\ell \nrep$ for $N_t=5,7$. Both cases show clear agreement between the two evaluation methods.}
    \label{fig:derivatives}
\end{figure}

Figure~\ref{fig:derivatives} shows the two quantities from relation \eqref{eq:d2F} as a function of $\mu$. 
For visual clarity, only two representative values of $N_t$ are displayed; the remaining values exhibit the same qualitative behavior. The excellent agreement between the results obtained from $\partial_{\ell}H_2$ and $\nrep$ confirms the internal consistency of the simulation algorithm. In addition, the results clearly resolve the finite-density phase transition at $\mu_c\approx 0.5$, illustrating that entanglement-based observables such as $\partial_{\ell}H_2$ or $\dSEE$ can be sensitive probes to investigate the phase structure of a theory.

To verify the validity of \eqref{eq:thermsdensmaxrel}, we recall that on the lattice, we use the approximation $\dSEE\approx\partial_{\ell}H_2$. We therefore have to focus on \eqref{eq:maxwell_rel_hr} with $r=2$, \ie, on
\begin{equation}
\frac{1}{V_{\perp}}\frac{\partial^2 H_2}{\partial\mu\,\partial\ell}\overset{{}_*}{=}-2\,N_t\,\of{n\of{2\,N_t,\mu}-n\of{N_t,\mu}}\ ,\label{eq:maxwell_rel_h2}
\end{equation}
instead of \eqref{eq:thermsdensmaxrel}, where the asterisk ($*$) 
refers to the requirement $\xi_{\text{max}}\ll\ell,N_x/2,N_s$ for the equality to hold.

Both sides of the relation \eqref{eq:maxwell_rel_h2} can be computed from lattice simulations. The left-hand side is obtained from $\partial_\mu\partial_{\ell}H_2$, while the right-hand side follows from measurements of the charge density \eqref{eq:lattice_n} in pairs of unreplicated systems of temporal extent $N_t$ and $2\,N_t$, respectively.

\begin{figure}[!htb]
    \centering
    \includegraphics[scale=0.375]{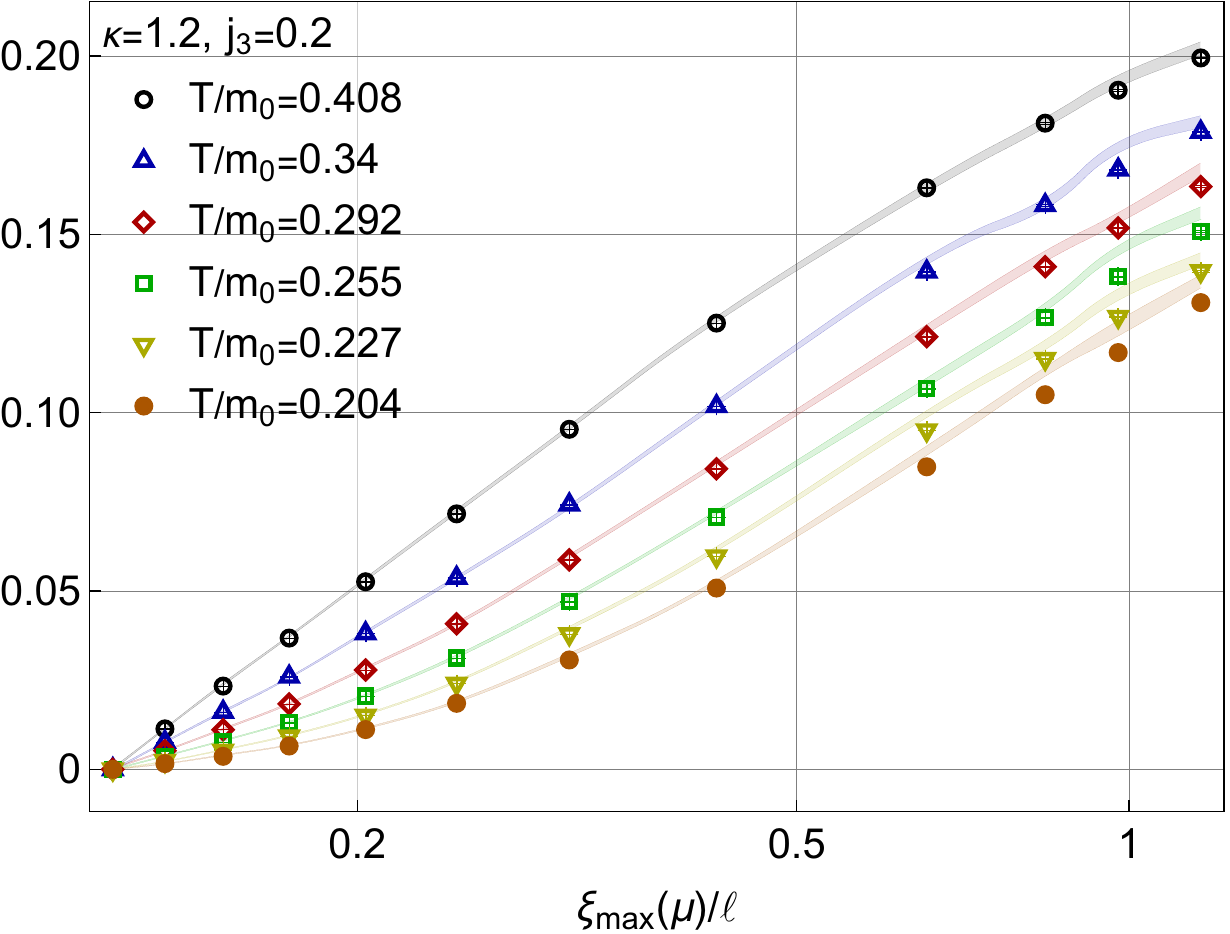}
    \caption{Comparison of the two sides of \eqref{eq:maxwell_rel_h2}. Bands show $V_\perp^{-1}\partial_\mu\partial_{\ell}H_2$ (evaluated via $\partial_\ell \nrep$), while points denote $-2\,N_t\,\of{n\of{2\,N_t,\mu}-n\of{N_t,\mu}}$, plotted as functions of $\xi_{\text{max}}\of{\mu}/\ell$. 
    Agreement persists up to $\xi_{\text{max}}\of{\mu}/\ell\approx 0.5$ at the lowest temperature and almost up to $\xi_{\text{max}}\of{\mu}/\ell\approx 1.0$ for the highest temperature.}
    \label{fig:maxwell_rel_h2}
\end{figure}

Figure~\ref{fig:maxwell_rel_h2} compares the two sides of \eqref{eq:maxwell_rel_h2}, showing $V_\perp^{-1}\partial_\mu\partial_{\ell}H_2$ (bands) and $-2\,N_t\,\of{n\of{2\,N_t}-n\of{N_t}}$ (point markers) as functions of $\xi_{\text{max}}(\mu)/\ell$. The displayed values for $V_\perp^{-1}\partial_\mu\partial_{\ell} H_2$ have been computed from measurements of $\partial_\ell\nrep$ by using \eqref{eq:d2F}, which yields a numerically slightly cleaner signal for $\mu<\mu_c$. As can be seen, the two quantities agree well up to $\xi_{\text{max}}/\ell\approx0.5$ for the lowest temperature and almost up to $\xi_{\text{max}}/\ell\approx1.0$ for the highest temperature. This can be understood, by noting that $\xi_{\text{max}}$ refers to the longest correlation length at zero temperature; at finite temperature, thermal effects truncate the effective correlation length.

In conclusion, we have argued on general grounds that the variation of EE with respect to the size of the entangling region approaches the thermal entropy density in the limit of large subregions. We further corroborated this relation through a nonperturbative lattice study of R\'enyi entropies in the three-dimensional $\On{4}$ model at finite density, where the relevant limits can be taken explicitly. Our results show that EE directly encodes thermodynamic response, including generalized Maxwell relations linking entropy, chemical potential, and charge density. Therefore, in interacting finite-density QFT,  entanglement variations can be treated on the same footing as thermodynamic response functions, rather than as purely information-theoretic diagnostics.

\textit{Acknowledgments}---This work was supported by the Research Council of Finland (grants 354533, 354572), the Centre of Excellence in Neutron-Star Physics (project 374062), the Finnish Quantum Flagship (project 358878), the Quantum Doctoral Education Pilot (QDOC VN/3137/2024-OKM-4), and the European Research Council (grant 101142449). Computational resources were provided by CSC -- IT Center for Science, Finland.

\bibliography{refs}

\end{document}